\documentclass[preprint,12pt]{elsarticle}

\usepackage{graphicx}
\usepackage{amssymb}
\usepackage{amsmath}

\journal{Journal of \LaTeX Templates}

\begin{document}
\begin{frontmatter}

\title{Goos-H\"anchen shifts due to 2D materials with complex conductivity}

\author{Ni\~{n}a Angelica F. Zambale$^{1}$, Jenny Lou B. Sagisi$^{1,2}$, and Nathaniel P. Hermosa$^1$}

\address{$^{1}$National Institute of Physics, University of the Philippines Diliman, Quezon City, Philippines}
\address{$^2$Material Science and Engineering Program, College of Science, University of the Philippines Diliman, Quezon City, Philippines}

\address{corresponding author: nazambale@nip.upd.edu.ph}

\begin{abstract}
We investigate theoretically the Goos-H\"{a}nchen (GH) shift of a $p$-polarized terahertz beam incident on a 2D material surface with complex conductivity. Taking monolayer graphene to be the model material, we determine the dependence of GH shifts on the Fermi level and incident frequency.  Both spatial and angular GH shifts are present. For both GH shifts in general, we find that increasing the Fermi level shifts the incident angle at which the maximum GH shifts arise.  Moreover, we see that at higher frequencies, the amount of beam shift decreases with the Fermi level when the incident frequency is changed. At lower frequencies, however, the shift becomes proportional with the Fermi level. Upon obtaining the measurable shifts, the angular GH shift dominates the spatial GH shift given appropriate experimental parameters. Our results may pave the way for these material's use in optoelectronics devices, and fundamentally, to determine properties of 2D materials with complex conductivity.
\end{abstract}


\end{frontmatter}

\section{Introduction}
A physical beam reflected on a surface of a material with an index gradient experiences a shift or deviation from the path predicted by geometrical optics. This effect is known as Goos-H\"{a}nchen (GH) shift \cite{goos1947neuer,aiello2012goos}. Because the GH shift can be related to the different material and structure parameters, measurement of this shift offers an alternative method to characterize properties of materials. Also with GH shifts, more materials could also be used as a photonics devices. Numerous material surfaces have been investigated for GH shifts such as metals \cite{leung2007large,merano2007observation}, dispersive media \cite{puri1986goos}, negatively refractive media \cite{berman2002goos}, and in metamaterials \cite{grzegorczyk2005reflection, lambrechts2016transformation, yallapragada2016observation}. However, the shift is generally extremely small that for practical applications, several models and experiments have been proposed to enhance the magnitude of shift either by using dispersive materials \cite{puri1986goos}, surface plasmon resonances \cite{cheng2014giant, yallapragada2016observation}, transformation optics \cite{lambrechts2016transformation}, beam shaping \cite{merano2010orbital,hermosa2012radial}, and weak measurements \cite{chen2017observation}.

Two-dimensional (2D) materials with complex conductivity recently gained attention after graphene, a 2D form of carbon exhibiting complex conductivity, was discovered in 2004 \cite{novoselov2004electric, geim2007rise}. Complex conductivity in materials arose from the Drude model of electrical conduction \cite{falkovsky2008optical,koppens2011graphene}. The imaginary factor in the Drude equation gives these materials unique properties important in electronics and optoelectronics applications where the imaginary conductivity plays a more significant role than the real part. 

Recently, a giant Goos-H\"{a}nchen shift was predicted and later experimentally observed in graphene under total internal reflection \cite{li2014experimental,cheng2014giant}. A large-scale lateral shift was observed  when the incident light changes from transverse magnetic (TM or $p$-polarized) to transverse electric (TE or $s$-polarized) mode. Furthermore, the tunability of the GH shifts due to graphene was theoretically calculated by Xu, et. al. wherein they presented the dependence of beamshifts on graphene\rq s chemical potential, and incident beam frequency and angle \cite{xu2016tunable}. 

Many studies have explored the optical properties of graphene in the visible region where graphene\rq s optical conductivity can be approximated as $\sigma = e^2 / 4\hbar$\cite{falkovsky2008optical,hermosa2016reflection, merano2016optical}. Graphene\rq s conductivity in the visible range is mainly contributed by the interband transition of carriers \cite{falkovsky2008optical}. However, when the frequency is at terahertz (THz), contributions from intraband scattering will dominate leaving the conductivity purely complex \cite{rouhi2012terahertz}. The conductivity of graphene in this range is seen to increase up to two orders of magnitude compared to the conductivity in the visible range. Moreover, the conductivity now depends on the free-carrier concentration. We can then control graphene\rq s optical and electronic properties in the terahertz range.

In this paper, we investigate theoretically the GH shift of a $p$-polarized terahertz beam due to graphene\rq s Fermi energy and incident frequency. The results presented here can aid in determining the optimal incident frequency and angle for a much easy detection of beamshift, which in turn, can be used to determine properties of materials. Furthermore, we show that by changing the Fermi energy by doping or potential gating, one could also control beam deflection for optoelectronic applications. We show that the shift is dominated by the angular GH shift whose effect are measurable with appropriate experimental parameters, while the spatial GH shift, although also easily measurable, is an order of magnitude smaller. Rarely any literature reports optimizing the GH shifts with 2D materials of complex conductivity with THz beam and with different $E_F$ in external reflection.

\section{Theoretical Framework}
\label{S:2}

\subsection{Fresnel coefficient.} 
The optical conductivity of graphene in the random phase approximation and at finite temperature $T$ can be written as \cite{koppens2011graphene,falkovsky2008optical},

\vspace{-.5cm}
\scriptsize
\begin{equation}
\sigma(\omega) = \frac{2e^2k_BT}{\pi\hbar^2}\frac{i}{\omega+i\tau^{-1}}\log \left[2\cosh \left(\frac{E_F}{2k_BT}\right)\right]+\frac{e^2}{4\hbar^2} \left[H \left(\frac{\omega}{2}\right)+\frac{4i\omega}{\pi} \int_0^\infty d\epsilon  \frac{H(\epsilon) - H(\frac{\omega}{2})}{\omega^2 - 4\epsilon^2} \right]
\label{eqn:con}
\end{equation}
\normalsize

\noindent where the first and second terms correspond to the intraband and interband transitions, respectively. This frequency dependence of the complex conductivity of homogeneous materials arises from the Drude model \cite{rouhi2012terahertz}. However, when the frequency goes down as in the THz range and the temperature $T$ is below $E_F/k_B$ which is usually the case at room temperature, the conductivity is reduced to,

\vspace{-0.6cm}
\small
\begin{equation}
\sigma (\omega) = \frac{2e^2}{\pi\hbar^2} \frac{iE_F}{\omega + i\tau^{-1}} 
\label{eqn:contrue}
\end{equation}
\normalsize

\noindent and becomes solely dependent on the scattering time $\tau$, Fermi level $E_F$ and frequency $\omega$ \cite{rouhi2012terahertz}. In this range, the interband transition is usually negligible due to the Pauli exclusion principle and as a result, the intraband scattering dominates the highly doped graphene\cite{cheng2014giant,rouhi2012terahertz}. For short-range scattering, the scattering rate $\tau$ is proportional to the Fermi level, $\tau = \mu E_f/e v_F^2$ where $\mu$ is the mobility and $v_F$ is the Fermi velocity\cite{li2017graphene}.  At room temperature, the mobility and the Fermi velocity have values in the order of $10000 \mathrm{cm}^2/\mathrm{Vs}$ and $1 \times 10^6 \mathrm{m/s}$, respectively\cite{fan2016electrically}. This conductivity affects light as it impinges on its interface. Moreover, when a 2D material is sandwiched between two dielectric surfaces with permittivity constants $\epsilon_1$ and $\epsilon_2$, the modified reflection coefficients are,

\vspace{-0.4cm}
\small
\begin{equation}
r_s = \frac{\sqrt{\epsilon_1}\cos\theta_1 - \sqrt{\epsilon_2}\cos\theta_2 - \tilde{\sigma}}{\sqrt{\epsilon_1}\cos\theta_1 + \sqrt{\epsilon_2}\cos\theta_2 + \tilde{\sigma}} \qquad \mathrm{and} \qquad
r_p = \frac{\frac{\sqrt{\epsilon_2}}{\cos\theta_2} -\frac{\sqrt{\epsilon_1}}{\cos\theta_1} +  \tilde{\sigma}}{\frac{\sqrt{\epsilon_1}}{\cos\theta_1} + \frac{\sqrt{\epsilon_2}}{\cos\theta_2} + \tilde{\sigma}},
\label{fresnel}
\end{equation}
\normalsize

\noindent where $\theta_1$ is the incident angle, $\theta_2$ is the transmitted angle, and $\tilde{\sigma}$ is the conductivity of the material as $\tilde{\sigma} = \frac{\sigma}{\epsilon_0\mathrm{c}}$. The Fresnel coefficients can be derived by imposing the boundary conditions: $ \mathbf{n \times (E_2 - E_1)|_{z=0} = 0}$ and $ \mathbf{n \times (H_2 - H_1)|_{z=0} = J}$ where \textbf{n} is the unit surface normal, $\mathbf{E_{1,2}}$ and $\mathbf{H_{1,2}}$ are the electric field and magnetic fields at the interface, respectively and \textbf{J} is the surface current density of material\cite{zhan2013transfer}.

\vspace{.15cm}
\noindent

\subsection{Goos-H\"anchen effect.} 
A real light beam with a finite beam width impinging on a plane dielectric interface exhibits spatial and angular shifts. These shifte, known as the Goos-H\"anchen shifts, rely on the polarization and frequency of the incident beam  and the material properties of the interface \cite{aiello2012goos,bliokh2013goos}. The dimensionless spatial ($\Delta_{GH}$) and angular ($\Theta_{GH})$ Goos-H\"anchen shifts are expressed as: 

\small
\vspace{-0.2cm}
\begin{equation}
\Delta_{GH} = w_p \textrm{Im} \left(\frac{\partial \ln r_p}{\partial \theta}\right) + w_s \textrm{Im} \left (\frac{\partial \ln r_s}{\partial \theta}\right)
\label{eqn:spa_GH}
\end{equation}
\vspace{-0cm}
\begin{equation}
- \Theta_{GH} = w_p \textrm{Re} \left(\frac{\partial \ln r_p}{\partial \theta}\right) + w_s \textrm{Re} \left (\frac{\partial \ln r_s}{\partial \theta}\right)
\label{eqn:ang_GH}
\end{equation}
\normalsize

\noindent where $w_{s/p} = \frac{{R^2}_{s/p}{a^2}_{s/p}}{R_p^2 a^2_p + R_s^2 a_s^2}$, ${a}_{s/p}$ are the electric field components of the incident beam, and ${r}_{s/p}$ are expressed in Eq. \ref{fresnel}\cite{aiello2012goos}. We used Eqs. \ref{eqn:spa_GH} and \ref{eqn:ang_GH} to calculate the shifts with Eqs. \ref{eqn:contrue} and \ref{fresnel} which are material dependent.

\section{Results and Discussion}
The conductivity of graphene in the THz range is dependent on several factors namely graphene\rq s Fermi level, incident beam frequency, and scattering time as seen in Eq. \ref{eqn:contrue}. In our calculations, a Polymethylpentene (TPX) substrate with permittivity of 2.1196 is used because it has a dielectric response to a THz beam. We assign values of the Fermi level from 0.1 eV to 0.2 eV which are experimentally realizable\cite{zouaghi2015good}. We have investigated the external reflection of a $p$-polarized THz beam for practicality and simplicity.

\begin{figure}[h!]
\centering
\includegraphics[width = \textwidth]{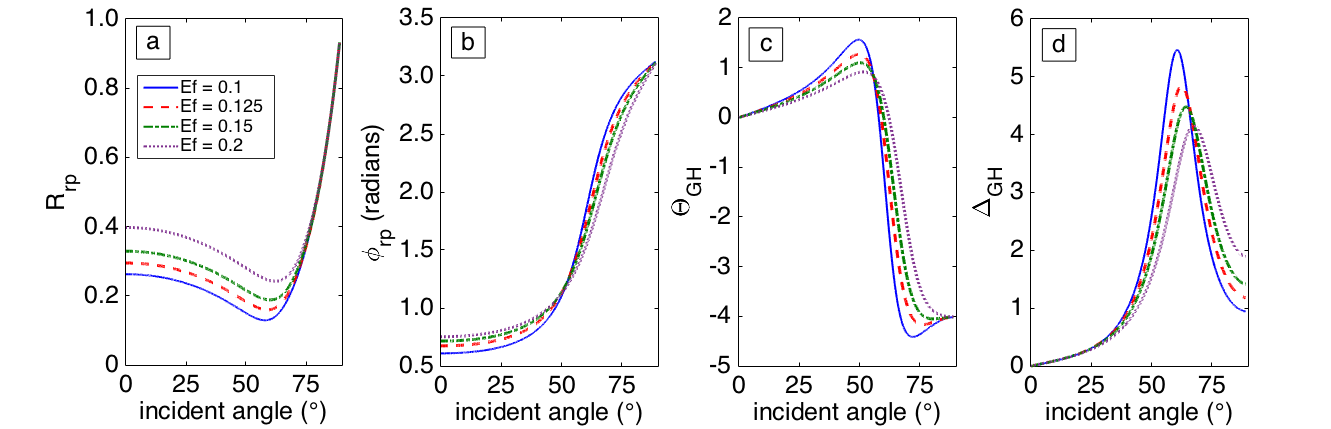}
\caption{Dependence of (a) Reflectivity and (b) phase of the beam, and the (c) angular and (d) spatial Goos H\"{a}nchen shifts on the incident angle for different Fermi levels ($\omega = 3$ THz). Increasing the Fermi level results to different values for the reflectivity and spatial Goos H\"{a}nchen shifts. With increasing $E_F$ and hence rising charge carrier concentration, the reflectivity becomes higher and its minima shift to larger incident angle while the angles where the inflection point of the phase happens also becomes higher. These result to a shift in the angle of incidence for maximum GH shifts.}
\label{fig:3THz}
\end{figure}

Figure \ref{fig:3THz} shows the reflectivity, phase jumps, and the Goos H\"{a}nchen shifts after the incident beam at an incident angle $\theta$ is reflected on an air-graphene-TPX interface. The reflectivity due to graphene never reaches zero: it attains a minimum value of reflectivity at certain incident angle referred to as the pseudo-Brewster angle $\theta_{pB}$. In general, the slope of each R$_{p}$ curve becomes steeper after passing $\theta_{pB}$. This behavior of reflectivity results to an angular GH shift, $\Theta_{GH}$ shown in Fig. \ref{fig:3THz}c. At lower incident angles, $\Theta_{GH}$ have increasing positive values and upon reaching $\theta_{pB}$, it decreases rapidly yielding negative angular Goos H\"{a}nchen shifts. By tuning the angle of incidence of the terahertz beam, one can change the polarity of the angular Goos H\"{a}nchen shift due to a monolayer graphene.

The phase shift of the THz beam as it reflects on the air-graphene-TPX interface is shown in Figure \ref{fig:3THz}b. For each Fermi level, $\phi_p$ grows steeper after passing a certain incident angle. There is no significant deviation between the phase shifts as the Fermi level is increased. However, a notable change can be observed in the spatial Goos H\"{a}nchen shift, $\Delta_{GH}$. The $\Delta_{GH}$ greatly rely on the incident angle of the THz beam. With increasing Fermi level, the $\Delta_{GH}$ decreases and its full width half max increases. Hence, by changing amount of charge carrier concentration of the monolayer graphene, the lateral shift of the incident THz beam can be changed also. This has significant implications in optoelectronic applications. 

\begin{figure}[h!]
\centering
\includegraphics[width = \textwidth]{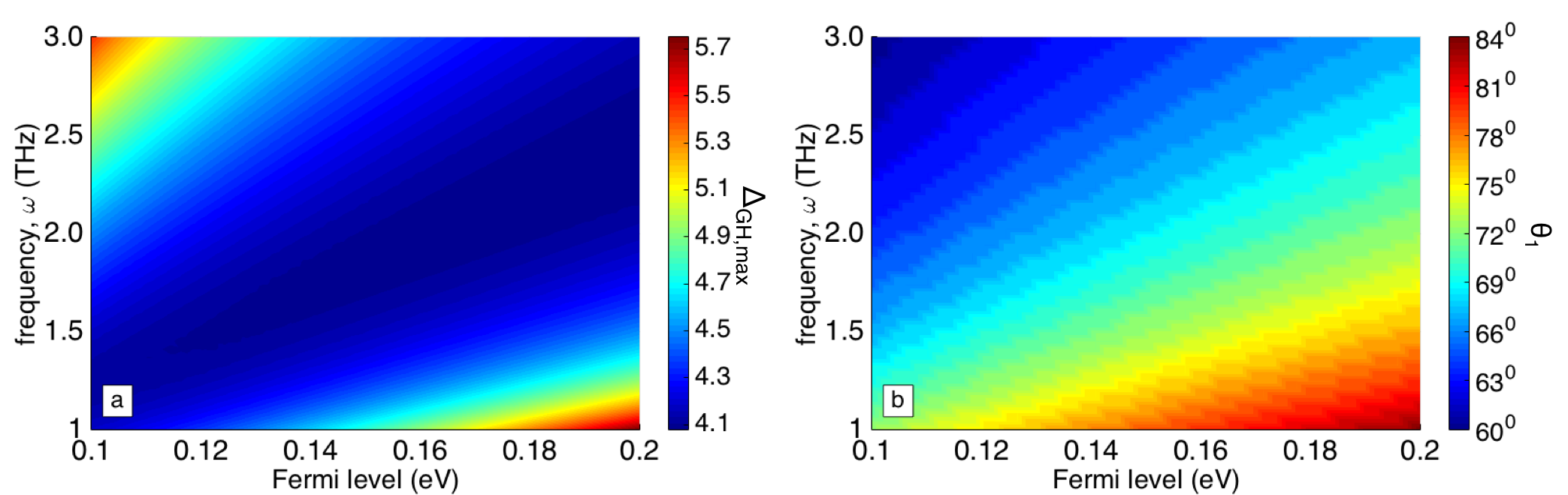}
\caption{Colormaps of (a) Maximum $\Delta_{GH}$ as a function of Fermi level and incident frequency and (b) the corresponding incident angles of each $\Delta_{GH,max}$. Large spatial Goos-H\"{a}nchen shifts can be obtained by having either low $E_F$ with high $\omega$, or high $E_F$ with low $\omega$. }
\label{fig:cmap_spaGH}
\end{figure}

The beamshifts are dependent on the frequency of the incident beam. To better visualize the dependence of the beamshifts with respect to both the Fermi level and the incident frequency, we have generated a color map of the maximum spatial shifts, $\Delta_{GH}$, and the incident angle $\theta$ where it occurs (Figure \ref{fig:cmap_spaGH}). At lower Fermi levels, the spatial GH shift peak grows as we increase the incident frequency. However, a different behavior is observed at higher Fermi levels where the increasing frequency decreases the $\Delta_{GH,max}$. Large $\Delta_{GH}$ can then be observed at the upper left and lower right corners which are regions of low Fermi levels with high incident frequencies and high Fermi levels with low incident frequencies, respectively. The values of the maximum shift ranges only from 4.0 to around 5.7,  corresponds to roughly $\sim \lambda/3$ to $\sim\lambda/2$, respectively, in measurable shifts. The $\Delta_{GH, max}$ do not significantly change. However, when we look at the corresponding angle where these large $\Delta_{GH}$'s occur, it is more practical to consider the shifts obtained from low Fermi levels with high incident frequencies because for the other region near normal incidence ($\theta\sim $84$^0$) is needed, which is quite difficult to achieve in experiments.

\begin{figure}[h!]
\centering
\includegraphics[width = \textwidth]{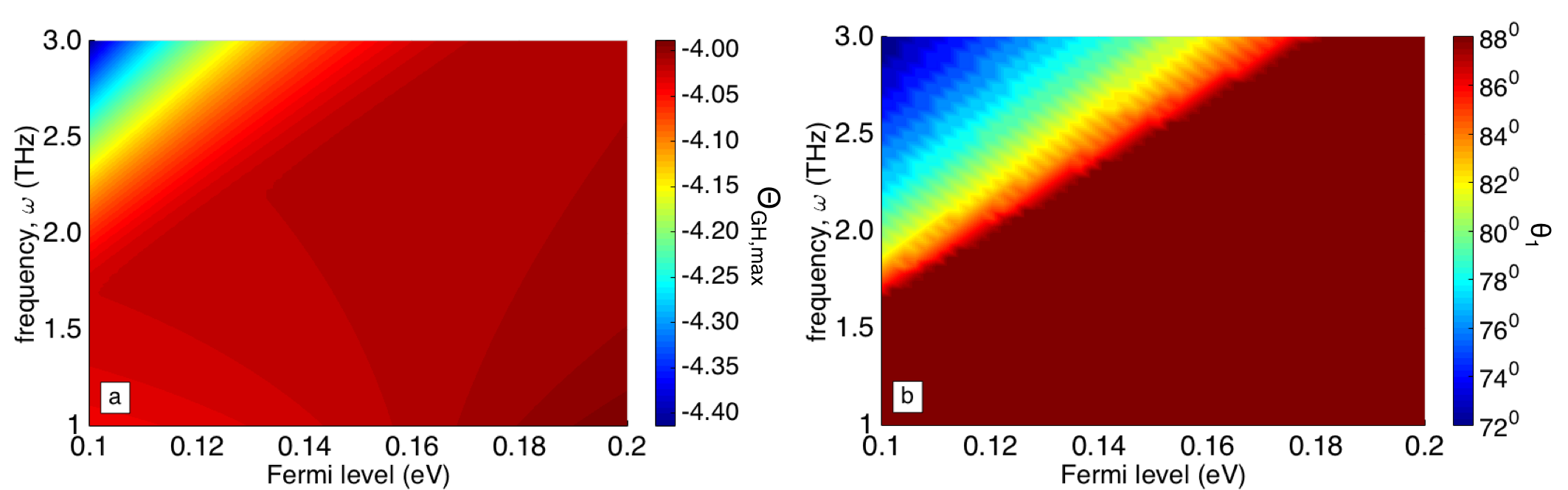}
\caption{Colormaps of (a) Maximum $\Theta_{GH}$ as a function of Fermi level and incident frequency and (b) the corresponding incident angles for each $\Theta_{GH, max}$. Large negative angular Goos-H\"{a}nchen shifts can be achieved with low $E_F$ and high $\omega$ values.}
\label{fig:cmap_angGH}
\end{figure}

The maximum angular GH shift $\Theta_{GH, max}$ color map shown in Fig. \ref{fig:cmap_angGH} is generated by taking to account the absolute maximum shift. The maximum angular shifts are negative. Similar to $\Delta_{GH}$, the most sensitive shift can be found at low Fermi levels with high incident frequencies. Again, the maximum shift do not vary greatly with values ranging from -4.0 to -4.4. Although, the measurable shift can be large as will be discussed below. At low incident frequencies, the maximum $\Theta_{GH}$ become independent of Fermi level. Maximum $\Theta_{GH}$ in each case can be achieved by setting the incident angle in accordance to Fig. \ref{fig:cmap_angGH}b. 

\begin{figure}[h!]
\centering
\includegraphics[width = \textwidth]{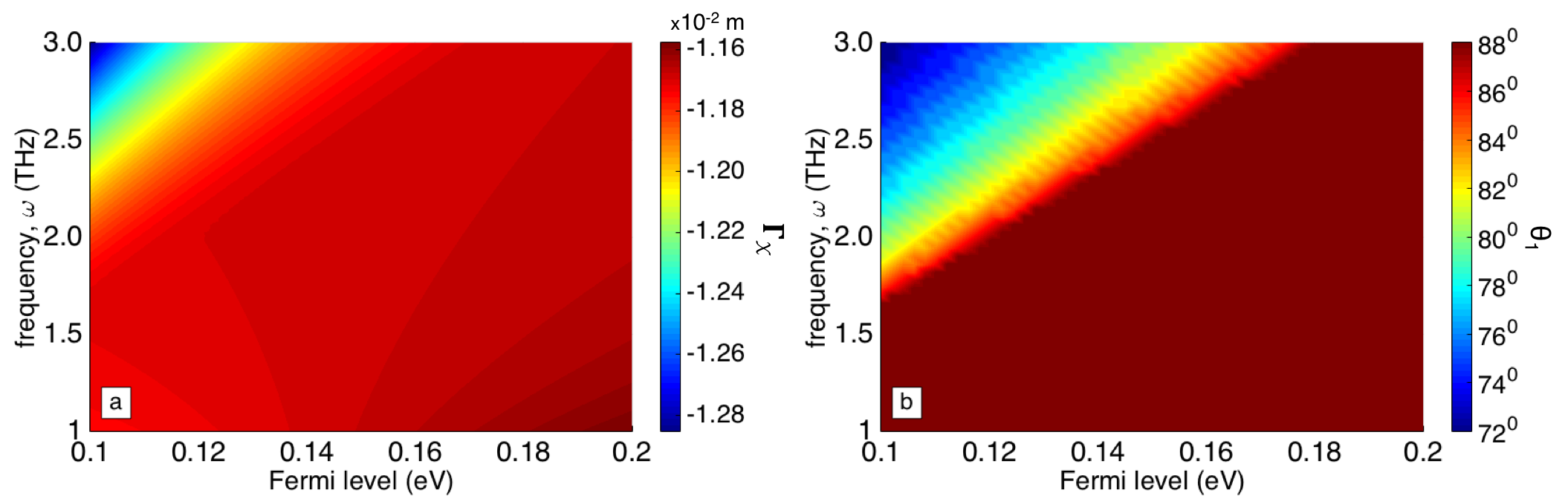}
\caption{Colormaps of (a) Maximum physical GH shifts as a function of Fermi level and incident frequency and (b) the corresponding incident angles for each maximum physical beamshift. We have calculated a physical beamshift of up to -13  mm (\emph{See text for details}.} 
\label{fig:cmap_phyGH}
\end{figure}

Both the dimensionless spatial and angular GH shifts discussed so far, are factors of how large the beamshift will be. It is useful therefore, to convert the these shifts into physical units that we can be measured in the laboratory. The physical beamshift $\Gamma_X$ is expressed as: $k_0\Gamma_X = \Delta_{GH} + (z/L) \Theta_{GH}$ where $k_0 = 2\pi/\lambda$, $z$ is the distance of the detector from minimum beamwaist, and $L = k_0\omega_0^2/2$ is the Rayleigh wavelength. We generated another colormap showing the behavior of the physical beamshift with different Fermi levels and incident frequencies (Figure \ref{fig:cmap_phyGH}). We have considered a detection distance of 23 cm from the focus of a lens and a THz beam with a beam waist of $\omega_0 = 2\lambda$. Notice that large physical shifts can be observed at low Fermi levels with high incident frequency, similar to what is observed for $\Theta_{GH, max}$ plot. This happens since $\Theta_{GH}$ dominates the $\Delta_{GH}$. At the specified propagation distance, we have calculated a physical beamshift up to $\approx -10^{-2}$ m occurring at around 70$^0$ incident angle. These shifts can be easily measured.

This work can be applied for characterizing 2D materials in terms of their Fermi level. In the case of graphene, one can measure either the minimum reflectivity (Figure \ref{fig:3THz}a) or the maximum $\Delta_{GH}$ (Figure \ref{fig:3THz}d) and $\Theta_{GH}$ (Figure \ref{fig:3THz}c). They have a one-to-one correspondence with the Fermi level of the material. Furthermore, we observe that maximum $\Delta_{GH}$ occurs near the pseudo Brewster angle. One can thus estimate the value of the Fermi level of a monolayer graphene by measuring either the reflectivity at the pseudo-Brewster angle or the peak of the spatial Goos-H\"{a}nchen shift. For easier detection of beamshift, $\Theta_{GH}$ can be used as the measurable shift can be modulated by a proper choice of lenses and detection distance. This opens up new techniques for non-invasive characterization of a 2D material.

\section{Summary and conclusions}
We showed the dependence of the Goos-H\"{a}nchen shift on the Fermi level of a 2D material with complex conductivity in the terahertz regime where graphene was chosen as the model material. We found that the Fermi level as well as the incident frequency can be tuned to maximize detection of Goos-H\"{a}nchen shift: low Fermi levels with high incident frequency favors large beamshifts. This can aid us in measuring the optical properties of graphene as well as other 2D materials with complex conductivity.

\section{Acknowledgement}
We would like to acknowledge the support of the University of the Philippines Office of the Vice President for Academic Affairs through the Balik PhD program (UP OVPAA 2015-06), University of the Philippines Enhanced Creative Work and Research Grant (ECWRG 2016-02-27), and the Philippine Council for Industry, Energy and Emerging Technology Research and Development (PCIEERD) of the Department of Science and Technology of the Republic of the Philippines.

\section*{References}
\bibliographystyle{model1-num-names}
\bibliography{sample.bib}

\end{document}